\documentclass[pra,aps,amsmath,amssymb,twocolumn,superscriptaddress]{revtex4}
\usepackage{times}
\usepackage{hyperref}
\usepackage{amsmath,amssymb}
\usepackage[usenames]{color}
\usepackage{grffile}
\usepackage{amsmath, amstext, amssymb, amsfonts, amsxtra}
\usepackage{textcomp}
\usepackage{xspace}
\DeclareSymbolFontAlphabet{\amsmathbb}{AMSb} 
\usepackage{soul}
\usepackage{xcolor}
\usepackage{epstopdf}
\usepackage{braket}
\usepackage{float}
\usepackage{tikz}
\usetikzlibrary{bending,arrows.meta,fadings}
\usepackage{pgfplots}
\usepgfplotslibrary{statistics}
\usepackage{enumitem}

\setlist{nosep}
\usetikzlibrary{arrows}

\newcommand{\bop}{\hat{b}} 

\newcommand{\nop}{\hat{n}}
 
\newcommand{\bdop}{\hat{b}^{\dagger}}

\newcommand{\rhop}{\hat{\rho}}

\newcommand{\tr}{{\rm tr}}

\newcommand{\Hop}{\hat{H}}

\newcommand{\smb}[1]{\left(#1\right)}

\newcommand{\pdt}[1]{\frac{\partial #1}{\partial t}}
\newcommand{\cmute}[2]{\left[{#1},{#2}\right]}
\newcommand{\ii}{\mathrm{i}}
\newcommand{\expe}{\mathrm{e}}

\newcommand{\Hs}{\Hop_{S}}

\newcommand{\drme}[2]{\mathcal{D}^{#1}\left[#2\right]}

\newcommand{\sutdepd}{EPD Pillar, Singapore University of Technology and Design, 8 Somapah Road, 487372 Singapore} 
\newcommand{\sutdsci}{Science and Math Cluster, Singapore University of Technology and Design, 8 Somapah Road, 487372 Singapore}

\begin{document}

\title{Interaction-impeded relaxation in the presence of finite-temperature baths}               

\author{Ryan Tan}     
\affiliation{\sutdepd}    
\author{Xiansong Xu} 
\affiliation{\sutdsci}
\author{Dario Poletti}
\affiliation{\sutdepd} 
\affiliation{\sutdsci}

\begin{abstract} 
We study the interplay between interactions and finite-temperature dephasing baths. We consider a double well with strongly interacting bosons coupled, via the density, to a bosonic bath. Such a system, when the bath has infinite temperature and instantaneous decay of correlations, relaxes with an emerging algebraic behavior with exponent $1/2$. Here we show that because of the finite-temperature baths and the choice of spectral densities, such an algebraic relaxation may occur for a shorter duration and the characteristic exponent can be lower than $1/2$. These results show that the interaction-induced impeding of relaxation is stronger and more complex when the bath has finite temperature and/or non-zero time scale for the decay of correlations. 
\end{abstract}

\maketitle

\section{Introduction}\label{sec:intro}      
The study of the effects of the environment on quantum systems is important for the development of future technologies. In recent years, a number of works have focused on various research directions. One example is the study of how one can engineer the bath to prepare interesting and useful quantum states \cite{DiehlZoller2008, KastoryanoSorensen2011, KienzlerHome2015, MetelmannClerk2015, KeckFazio2018, CianHafezi2019}. Another example is the study of how one can protect the system from the environment, for instance using external factors, e.g. drivings \cite{ViolaLloyd1999}, or due to intrinsic properties of the system which make it more resistant against the bath \cite{PolettiKollath2012, PolettiKollath2013, SciollaKollath2015, BernierKollath2018}. In Ref. \cite{PolettiKollath2012, CaiBarthel2012, PolettiKollath2013, SciollaKollath2015, BernierKollath2013, RenWang2019, WuEckardt2019}, it was stressed that interesting physics emerges not only in the steady state, but also in the transient dynamics towards the steady state. In particular, in Ref. \cite{PolettiKollath2012}, it was shown that a strongly interacting double-well system with enough bosons shows a power-law relaxation regime when coupled to a bath via dephasing, a phenomenon referred to as {\it interaction-induced impeding}. In Ref. \cite{PolettiKollath2013}, the authors also showed the emergence of a stretched exponential regime, i.e. the relaxation dynamics has a time-dependence of the type $\exp(-(t/\tau)^{\alpha_s})$ where $\alpha_s$ is an exponent between $0$ and $1$ and $\tau$ is a time scale. Furthermore, in \cite{SciollaKollath2015} they showed the emergence of aging \cite{HenkelSanctuary2007}, i.e. the two-time correlation between two times $t_1$ and $t_2$ does not depend on their difference, $t_2-t_1$, but is proportional to $(t_2/t_1)^{\alpha_a}$, where $\alpha_a$ is another exponent. These results were obtained both numerically and analytically. In particular, by using the ``adiabatic-elimination'' technique \citep{GarciaRipollCirac2009, PolettiKollath2012}, one could show that these dynamics can be described by classical anomalous diffusion equations.  
Since then, the adiabatic elimination approach has been used to gain deeper insights for various many-body open quantum system \cite{PolettiKollath2013, BernierKollath2013, BernierKollath2014, SciollaKollath2015, MedvedyevaZnidaric2016, BernierKollath2018, WolffKollath2019}. Complex relaxation dynamics have also been observed in dissipative Rydberg atoms setups which can be modeled by systems with kinetic constraints, both in theory \cite{OlmosGarrahan2012, LesanovskyGarrahan2013} and experiments \cite{MalossiMorsch2014, ValadoMorsch2016, GutierrezMorsch2017}.  

Previous works studying the effect of dephasing on many-body quantum systems relied on the fact that the master equation would drive the system towards the infinite temperature state \cite{PolettiKollath2012, PolettiKollath2013, SciollaKollath2015, BernierKollath2018, WolffKollath2019}. Hence, for however weak the coupling between the system and the bath was, it would always be possible to eventually heat up the system such that all the possible configurations of atoms or spins would become equally probable. Considering a strongly interacting system with many particles, for this infinite-temperature approach to be valid, one would need temperatures larger than any energy difference in the system.
It is thus important to study how such relaxation dynamics would be affected by considering finite-temperature baths, and also examine the effect of structured baths. That is the aim of this work. In the following we will show that at finite temperatures and/or for spectral densities with a finite range of frequencies, the duration of the algebraic regime can be significantly shortened, and the exponent may change to a smaller value. For the description of the effects of the environment on the system we have used a Redfield master equation \cite{Redfield1957}, which was recently also used to tackle many-body quantum systems \cite{XuPoletti2019}, showing accurate dynamics as compared to numerical exact approaches \cite{deVegaBanuls2015, GuoPoletti2018}.

This manuscript is structured in the following way: In Sec. \ref{sec:model} we describe in detail the studied model, with particular emphasis on how we consider the effects of dissipation in Sec. \ref{ssec:bath}. In Sec. \ref{sec:results} we present our results of the relaxation dynamics of the system as a function of the temperature of the bath and its spectral properties. In Sec. \ref{sec:conclusions} we draw our conclusions.

\section{Model}\label{sec:model} 
We study a deep double-well potential in which each site is coupled to a finite-temperature bosonic bath. We take the total Hamiltonian $\Hop_{\rm tot}$, which includes both the system and bath, to be time independent, and in particular to have the generic form  
\begin{align}  
	\Hop_{\rm tot} = \Hop_{S} + \sum_i (\Hop_{B_i} + \hat{S}_i\hat{B}_i),
\end{align} 
where $\Hop_S$ is the Hamiltonian of the system under consideration, $\Hop_{B_i}$ is the $i$th-bath Hamiltonian, and the interaction between system and each bath is given by $ \hat{S}_i \hat{B}_i$ where $S_i$ acts on the system while $\hat{B}_i$ acts on the $i$th bath. We consider the case in which the system-bath coupling is weak, and the initial global density matrix of the system and bath $\rhop_{\rm tot}(0)$ is in a separable form $\rhop_{\rm tot}(0)\approx \rhop(0)\bigotimes^2_{i=1} \rhop_{B_i}$, where the reduced density matrix $\rhop(0)$ describes the system while $\rhop_{B_i}$ is a thermal Gibbs state for the $i$th bath at temperature $T$. By the Born approximation and time-local Markov approximation \cite{Breuer2010, deVegaAlonso2017}, the evolution of $\rhop(t)$ is given by   
\begin{align}
	\pdt{\rhop\smb{t}}=&-{\ii}\cmute{\Hs}{\rhop\smb{t}}+\drme{t}{\rhop{\smb{t}}},   
	\label{eq:rme}
\end{align}
which is also known as the Redfield master equation (RME) \cite{Redfield1957, deVegaAlonso2017}. In Eq. (\ref{eq:rme}) the first term on the right-hand side describes the unitary evolution due to the system Hamiltonian, while the dissipation due to the baths is described by a time-dependent superoperator 
\begin{align}
	\label{eq:relaxation}
	\drme{t}{~\cdot~}=& \sum_i \cmute{\mathbb{S}_i\smb{t}~\cdot~}{\hat{S}_i}+\cmute{\hat{S}_i}{~\cdot~\mathbb{S}_i^\dagger\smb{t} }, \\
	\mathbb{S}_i\smb{t}=&\int^{t}_0 \tilde{S}_i\smb{-\tau}C_i\smb{\tau} d\tau,
	\label{eq:transition}
\end{align}  
with $\tilde{S}_i\smb{\tau}=\expe^{\ii\Hs\tau}\hat{S}_i\expe^{-\ii\Hs\tau}$ while the bath correlation function is 
$C_i\smb{\tau}={\rm tr}\!\left(e^{\ii \Hop_{B_i} \tau} \hat{B}_i e^{-\ii \Hop_{B_i} \tau} \; \hat{B}_i \;\rhop_{B_i}\right)$ 
and 
$\rhop_{B_i}=e^{-\Hop_{B_i}/T}/\tr\left(e^{-\Hop_{B_i}/T}\right)$.
Note that we use units such that $\hbar=k_B=1$, where $k_B$ is the Boltzmann constant.  
Here we refer to $\mathbb{S}(t)$ as the transition operator which carries all the relevant bath information, including its temperature and spectral properties.  
In the following we will use $\mathbb{S}(\infty)$ in our computations because we tested that the dynamics of the system is identical, for the duration of the algebraic relaxation regime, to the one obtained using $\mathbb{S}(t)$. This approximation makes the Redfield master equation Markovian in the sense that the first-order evolution of the density operator $\rhop(t)$ does not effectively depend on the previous times.

Here we should discuss briefly one important aspect of the Redfield master equation, which is that it does not guarantee that the density matrix remains non-negative \cite{IshizakiFleming2009, Breuer2010}. We have thus checked this issue by computing the eigenvalues $\lambda_i$ of the density matrix representing the system, added together only the negative ones, and considering the absolute value of this sum $\epsilon = \sum_{i\;{\rm for}\;\lambda_i<0} |\lambda_i|$. We found that $\epsilon$ is negligible \cite{fn2}.

\subsection{The system Hamiltonian}\label{ssec:hamiltonian} 
We consider bosons in a double-well potential with the Hamiltonian 
\begin{align}
\hat{H}_S = - J \left( \bdop_1\bop_2^{} + \bdop_2\bop_1^{} \right) + \frac{U}{2} \left[\nop_1(\nop_1-1) + \nop_2(\nop_2-1)\right] \label{eq:Ham}             
\end{align}   
where $J$ is the tunneling amplitude, $U$ is the interaction magnitude, $\bdop_j$ ($\bop_j^{}$) creates (destroys) a boson at site $j$, $\nop_j=\bdop_j\bop_j$ and $[\bop_i^{}, \bdop_j] = \delta_{i,j}$.   
For this model, we are interested in the regime of both strong interaction $U$ and large particle number $N=\sum_{j=1,2} \tr(\rho \nop_j)$. The interactions are needed to explore the impeded regime, in which the interactions slow down the relaxation dynamics, 
while the large particle number makes the emergence of the algebraic regime possible, instead of simple exponential decays.       

\subsubsection{Initial conditions}\label{sssec:initialcondition}     

In the regime of large $U$ and total particle number $N$, it is possible to approximate the Hamiltonian of the system as that of a harmonic oscillator. In fact, we can write the state of the system with a basis $| n_1 \rangle$ with $N+1$ elements, where $n_1$ is the number of atoms in well $1$. Hence, for large $N$ the kinetic component of the Hamiltonian (the portion multiplied by $J$) is approximately constant near $n_1=N/2$ (using $\sqrt{N/2+1}\approx\sqrt{N/2}$), while the interaction term is proportional to $(n_1-N/2)^2$. Near $n_1\approx N/2$, we can thus approximate the Hamiltonian given by Eq.(\ref{eq:Ham}) with a single-particle Hamiltonian $\Hop_{sp}$, 
\begin{align} 
	\Hop_{sp} = -2JN^3 \frac{\partial^2}{\partial x^2}+ U N^2 \hat{x}^2 \label{eq:Hams}
\end{align}
where $\hat{x}=(\hat{n}_1-N/2)/N$. For $U$ large enough, the ground state is, with much higher probability, close to $x=0$ and thus the approximation of the kinetic term is accurate. From Eq. (\ref{eq:Hams}) we get that the ground state is a Gaussian with the variance proportional $\sqrt{JN/U}$. We thus expect the particle fluctuations in the first site, $\kappa(t=0) = \langle\nop_1^2\rangle - \langle\nop_1\rangle^2$, to scale as $\sqrt{N}$ for the ground state.    

\subsection{Modeling of the bath}\label{ssec:bath}  

We consider a structured environments with a finite temperature modeled by a collection of harmonic oscillators. The bath Hamiltonian is given by   
\begin{align}   
\Hop_{B_i}=\sum^\infty_{n=1}\left[\frac{\hat{p}_{i,n}^2}{2m_{i,n}}+\frac{m_{i,n}\omega_{i,n}^2 \hat{x}_{i,n}^2}{2}\right]. 
\end{align} 

Since we aim to study dephasing, we consider a coupling with the system operator given by $\hat{S}_i=\nop_i$ and 
$\hat{B}_i=-\sum^\infty_{n=1}c_{i,n} \hat{x}_{i,n}$ 
where 
$c_{i,n}$ 
is the coupling constant for the $n$th mode of the $i$th bath. 
From the definition of $C(\tau)$, we get that 
$C_i(\tau)=\int_0^\infty \left(d\omega/\pi\right){\rm J}_i(\omega) \left[\coth\left(\omega/2T\right)\cos(\omega\tau) - \ii \sin(\omega\tau)\right]$, with ${\rm J}_i(\omega)=\pi \sum c_{i,n}^2/(2m_{i,n}\omega_{i,n})\delta(\omega-\omega_{i,n})$. Here we note that this system-bath coupling results in nonquadratic equations even for $U=0$.     

The bath properties can be characterized by the spectral function ${\rm J}_i(\omega)$ which we consider to be ohmic with an exponential suppression depending on a cut-off frequency $\omega_c$, i.e., ${\rm J}_i(\omega)=\gamma_i \omega \exp(-\omega/\omega_c)$, and where $\gamma_i \propto\sum_n c_{i,n}^2 $ is the dissipation strength. In the following we consider the baths to be identical and coupled with the same magnitude to the system $\gamma_i=\gamma$, except that one bath is coupled with the density to the first site, while the other bath is coupled to the second site.    
The spectral function describes the energy differences present in the bath and which can cause excitations in the system. As we shall see later, the interplay between the energies available in the bath and the energy differences in the system has an important role in the relaxation dynamics.  


To better understand the role of the bath, we study the transition operator $\mathbb{S}(t)$ in Eq. (4) in the eigenbasis of the system Hamiltonian. In long-time limit $t \rightarrow \infty$, the real part of the transition operator $\mathbb{S}(\infty)$ can be found analytically, i.e., $Re(\Bra{n} \mathbb{S}(\infty) \Ket{m}) =  S_{n,m}W_{n,m}$, where $\Ket{n}$ represents the $n$th eigenstate of the Hamiltonian. The $W_{n,m}$ are often referred to as the transition rates, while $S_{n,m}$ are the elements of the system operator $\hat{S}$ in the eigenbasis of the system Hamiltonian.       
Focusing on the off-diagonal rates, the transition rates are given by (see, e.g., \cite{WichterichMichel2007, Breuer2010, ThingnaHanggi2013})  
\begin{align}
\label{eq:rates}
\begin{split}
W_{n,m}& = {\rm J}(\Delta_{n,m}) \; n_B(\Delta_{n,m}, T), \quad  \Delta_{n,m} > 0,
\\
W_{n,m}& = -{\rm J}(-\Delta_{n,m} \;) n_B(-\Delta_{n,m}, T), \quad  \Delta_{n,m} < 0.
     \end{split}
\end{align}
where $\Delta_{n,m}=E_n-E_m$ is the level spacing between levels $n$ and $m$, and $n_B(\Delta_{n,m}, T)$ is the Bose-Einstein statistics.

\subsection{Infinite temperature local Lindblad bath}\label{ssec:revision}   
To help the reader, here we review the dynamics of the interacting double well coupled to an infinite temperature Markovian bath which can be modeled by a local master equation in Gorini-Kossakowski-Sudarshan-Lindblad form \cite{Lindblad1976, GoriniSudarshan1976}, 
\begin{align}
	\pdt{\rhop\smb{t}}=&-{\ii}\cmute{\Hop_S}{\rhop\smb{t}}+\Gamma \sum_i [\nop_i, [\rho, \nop_i]],   
	\label{eq:lme}
\end{align}   
where $\Gamma$ is a dissipation rate. This master equation was derived for ultracold atoms under the effect of spontaneous emissions in \cite{PichlerZoller2010, GerbierCastin2010}. For the setup described by Eq. (\ref{eq:lme}), it was shown that if $U/J$ and $N$ are large enough, then (i) the dynamics would depend on a new timescale given by $U^2N^2/(\Gamma J^2)$, and (ii) if the initial condition is a balanced enough occupation of the two wells, then the probability distribution of the number of particles in one of the wells would be accurately described by an anomalous diffusion equation \cite{PolettiKollath2012}.    
Such anomalous diffusion could be measured by the time dependence of the particle fluctuation number $\kappa = \langle\nop_i^2\rangle - \langle\nop_i\rangle^2$, which would show a power law regime proportional to $\sqrt{t}$. An identical power-law has also been predicted in Ref. \cite{PolettiKollath2012} for the coherence, and in a later work these results have been extended to larger systems \cite{PolettiKollath2013, SciollaKollath2015}. This anomalous diffusion behavior has recently been tested experimentally in \cite{BouganneGerbier2019}.  
However, 
both the temperature and the spectral density could significantly change the emerging dynamics, whose study is the main goal of our investigation.

\section{Results}\label{sec:results} 
\begin{figure}[t]
	\centering
	\includegraphics[width=\columnwidth]{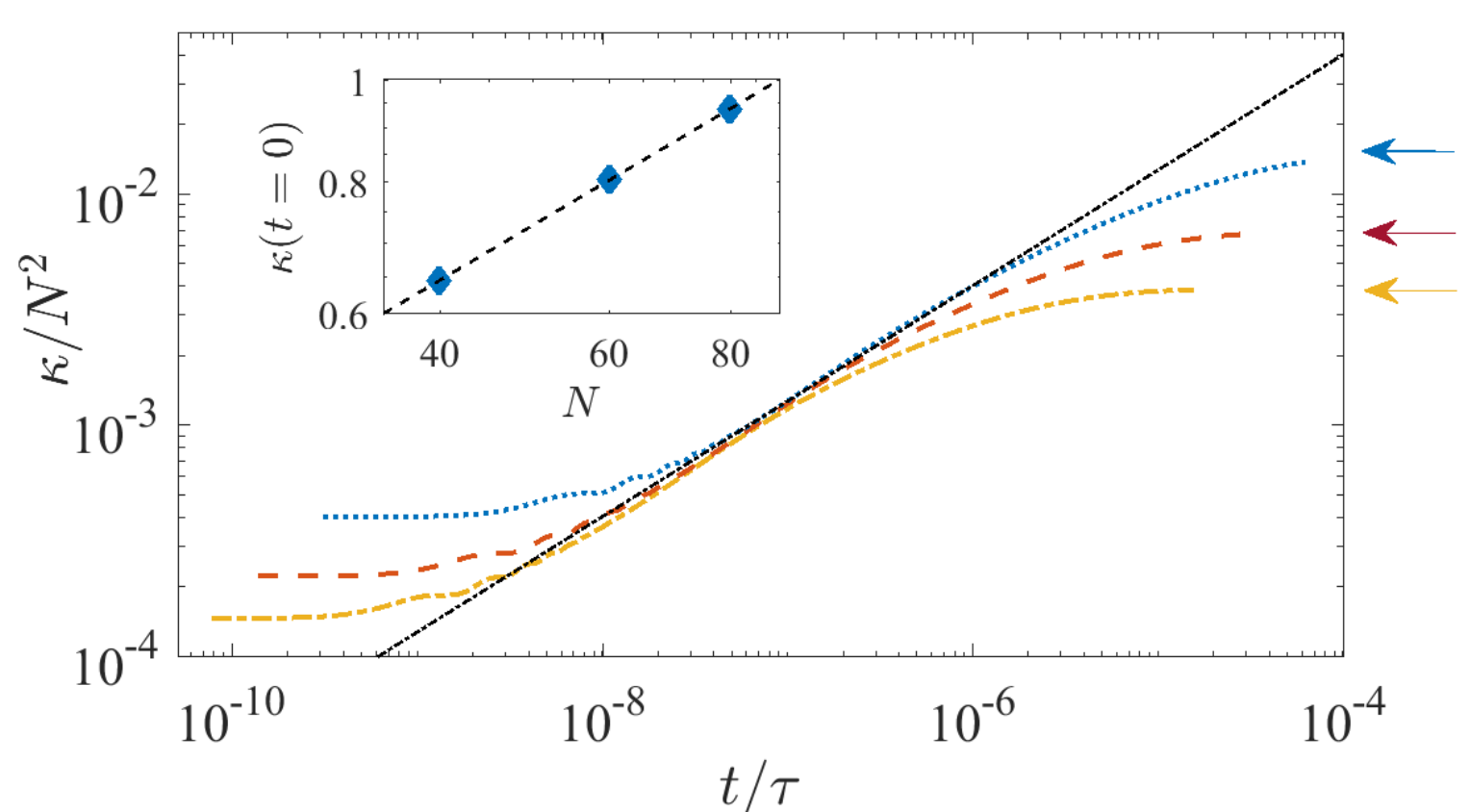}
	\caption{ Rescaled density fluctuations $\kappa/N^2$ vs rescaled time $t/\tau$ for different total particle numbers $N = 40\text{ (blue dotted line)}$, $60\text{ (red dashed line)}$ and $80\text{ (yellow dash-dotted line)}$, with temperature $T/J = 500$ and cut-off frequency $\omega_c/J =500$. The arrows point to the asymptotic values of the rescaled fluctuations. Inset: Scaling of $\kappa$ with $N$ for the initial condition, $\kappa \sim \sqrt{N}$ as predicted by the harmonic approximation, given by Eq.(\ref{eq:Hams}).}
	\label{fig:Fig2}
\end{figure} 

We now consider the relaxation dynamics of the system with the Hamiltonian given by Eq. (\ref{eq:Ham}) following the master equation from Eq. (\ref{eq:rme}) due to the finite-temperature baths. In the following, we will study the effects of temperature $T$, interaction strength $U$, particle number $N$, and the cut-off frequency $\omega_c$. We stress that for an infinite temperature bath, the steady state value of the fluctuation $\kappa/N^2$ is independent of $U$ and is given by $1/12+1/6N$. This also means that for large enough $N$, the steady state $\kappa/N^2$ converges to $\kappa/N^2\rightarrow 1/12$, which is independent of $N$. However, for finite temperature, the steady-state value of $\kappa/N^2$ is dependent on both $U$ and $N$. It is thus natural to expect that the power-law regime that emerges from the interplay between the interaction, dissipation, and kinetic energy, would survive for a shorter duration, which decreases as the interaction $U$ or $N$ increases. At the same time, if the temperature is too small, then the power-law regime may not even occur or it may be following a different exponent. This is what we are going to analyze in the following.  

We first consider the effect of different interaction strengths and the particle number $N$ on the dynamics when the temperature and frequency cutoff are large. 
In Fig. \ref{fig:Fig2} we analyze the effect of the particle number while keeping the interaction constant at $U/J=10$. Here we plot the rescaled particle fluctuations $\kappa/N^2$ versus rescaled time $t/\tau$ where $\tau=U^2N^2/(\gamma J^2)$. The blue dotted curve depicts the fluctuations for $N=40$, the red dashed one for $N=60$ and the yellow dash-dotted one for $N=80$.  
In Fig. \ref{fig:Fig2} we observe that $\kappa/N^2$ is dependent of $N$, unlike for the infinite temperature case. In particular, the arrows in Fig. \ref{fig:Fig2} indicate the asymptotic values of the particle fluctuations. Because of the different steady-state values, given a $U/J$ and a temperature $T$, the evolution of the rescaled fluctuations will deviate from the power-law curve at earlier rescaled times. Despite this, we still observe that there is a good enough merging of the curves and they share the same fit to a power-law curve with exponent $1/2$, as predicted in Ref. \cite{PolettiKollath2012}.      
In the inset of Fig. \ref{fig:Fig2}, we depict the dependence of the initial time fluctuations (i.e., the ground-state fluctuations) for the double well, showing the power-law regime predicted to be $\kappa \propto \sqrt{N}$ in Sec.\ref{sssec:initialcondition}.

\begin{figure}[t]
	\centering
	\includegraphics[width=\columnwidth]{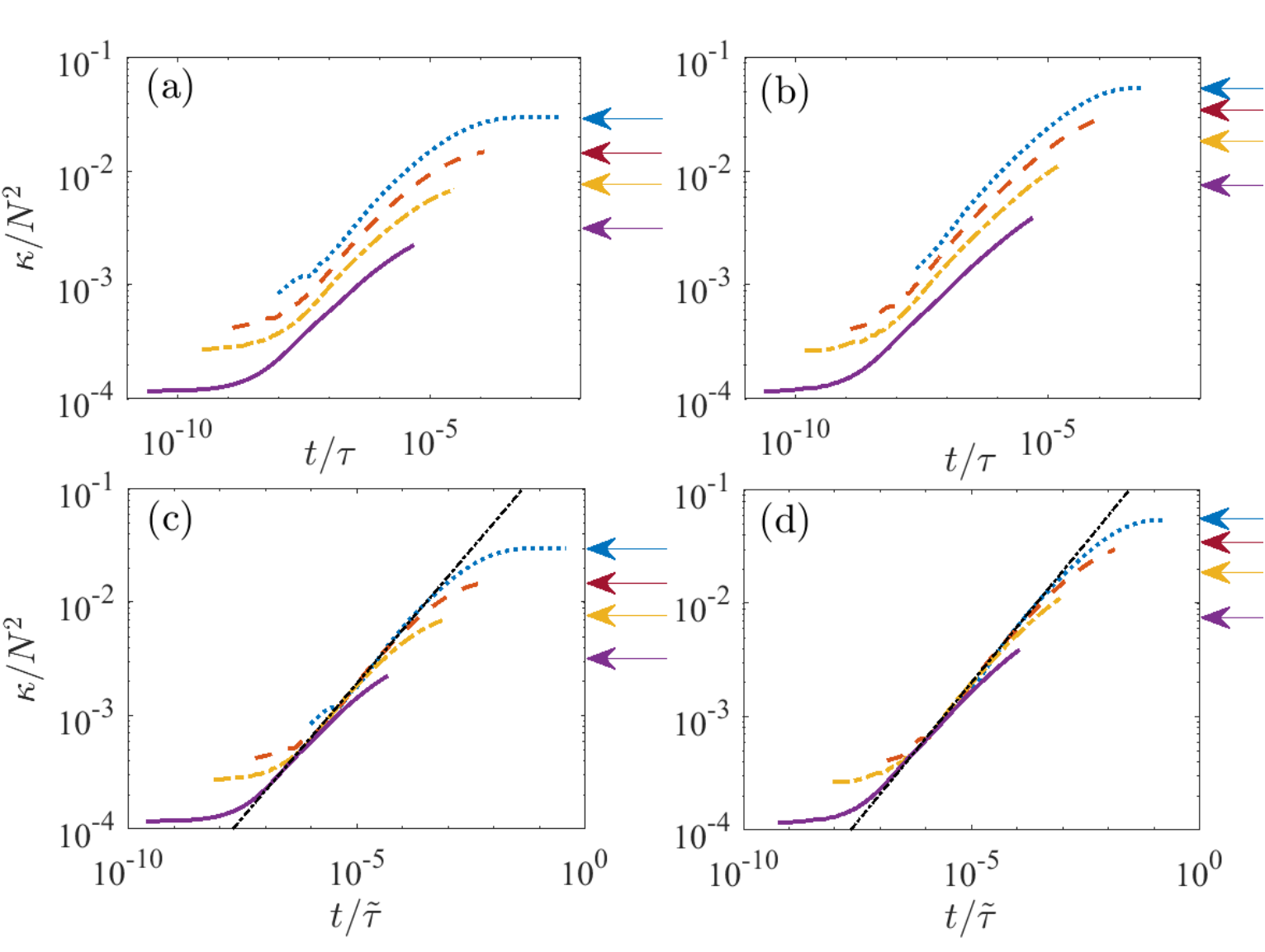}
	\caption{Density fluctuations $\kappa/N^2$ vs (a),(b) rescaled time $t/\tau$ or vs (c),(d) relaxation-rate-adjusted rescaled time $t/\tilde{\tau}$, for different interaction strengths $U/J = 5$ (blue dotted line), $10$ (red dashed line), $20$ (yellow dash-dotted line), and $50$ (purple continuous line). In (a),(c), the bath temperature is $T/J = 500$, while in (b),(d), $T/J = 1200$. The cutoff frequency $\omega_c/J = 500$ and $N=40$. The black dashed line shows the power-law regime with exponent $1/2$. }
	\label{fig:Fig1}
\end{figure}

In Figs. \ref{fig:Fig1}(a) and \ref{fig:Fig1}(b), we show the rescaled fluctuation $\kappa/N^2$ versus the rescaled time $t/\tau$, with different temperatures $T=500J$ [Fig.~\ref{fig:Fig1}(a)] and $T=1200J$ [Fig.~\ref{fig:Fig1}(b)] being used. The different lines represent different interaction strength $U$; in particular, the blue continuous line is for $U/J=5$, the red dashed line is for $U/J=10$, and the green dot-dashed line is for $U/J=20$. The dotted black line shows the power-law regime predicted for an infinite-temperature bath. The steady-state value of $\kappa/N^2$, for the values studied in Fig. \ref{fig:Fig1}, is given by the arrows of the colors of the corresponding lines.
As expected, we observe in Fig. \ref{fig:Fig1} that the curves for larger interactions deviate from the power-law line at earlier rescaled times. By comparing Figs.~\ref{fig:Fig1}(a) and \ref{fig:Fig1}(b), we also observe that for larger temperatures, the curves of the fluctuation follow the power-law line for longer rescaled times, as the baths are able to bring the system to larger values of $\kappa$.

Unlike Fig. \ref{fig:Fig2}, in Figs. \ref{fig:Fig1}(a) and \ref{fig:Fig1}(b) there is no clear collapse of the curves when we plot the fluctuations versus the rescaled time $t/\tau$. This is due to the fact that the relaxation rates $W_{n-1,n}$ given by Eq. (\ref{eq:rates}), which couple two consecutive eigenstates of the Hamiltonian, depend on the energy difference between these eigenstates. For a given large interaction, the energy gaps between the eigenstates, $E_n-E_{n-1}$, grow approximately as a linear function of the energy level $n$. This linear function is independent of $N$, but proportional to the interaction $U$. For this reason, there is a good scaling and collapse when varying $N$ at fixed $U/J$, but not when varying $U/J$ at fixed $N$. For large-enough temperatures and $\omega_c$, we realize that we can use a different time scale, rescaled with $W_{0,1}$, to obtain a better collapse of the particle fluctuations. This is shown in Figs. \ref{fig:Fig1}(c) and \ref{fig:Fig1}(d) in which we plot $\kappa/N^2$ versus $t/\tilde{\tau}$, where $\tilde{\tau}=\tau/W_{0,1}$, and where the curves represent the fluctuations for the corresponding values of $U$ as in Figs. \ref{fig:Fig1}(a) and \ref{fig:Fig1}(b). The corrected timescale $\tilde{\tau}$ allows one to fit the power-law regime for all the curves with the same algebraic line, $\propto \sqrt{t}$. When comparing Figs.~\ref{fig:Fig1}(c) and \ref{fig:Fig1}(d), it is clear that higher temperatures allow the power-law regime to last for a longer time. 
We note, however, that this is possible because the temperature is fairly large and so is the cutoff $\omega_c$, and hence the various $W_{n-1,n}$ do not vary significantly for a sizable number of energy levels. However, this will not be the case any longer for small $T$ and $\omega_c$, which we investigate in the following.

\begin{figure}[t]
	\centering
	\includegraphics[width=\columnwidth]{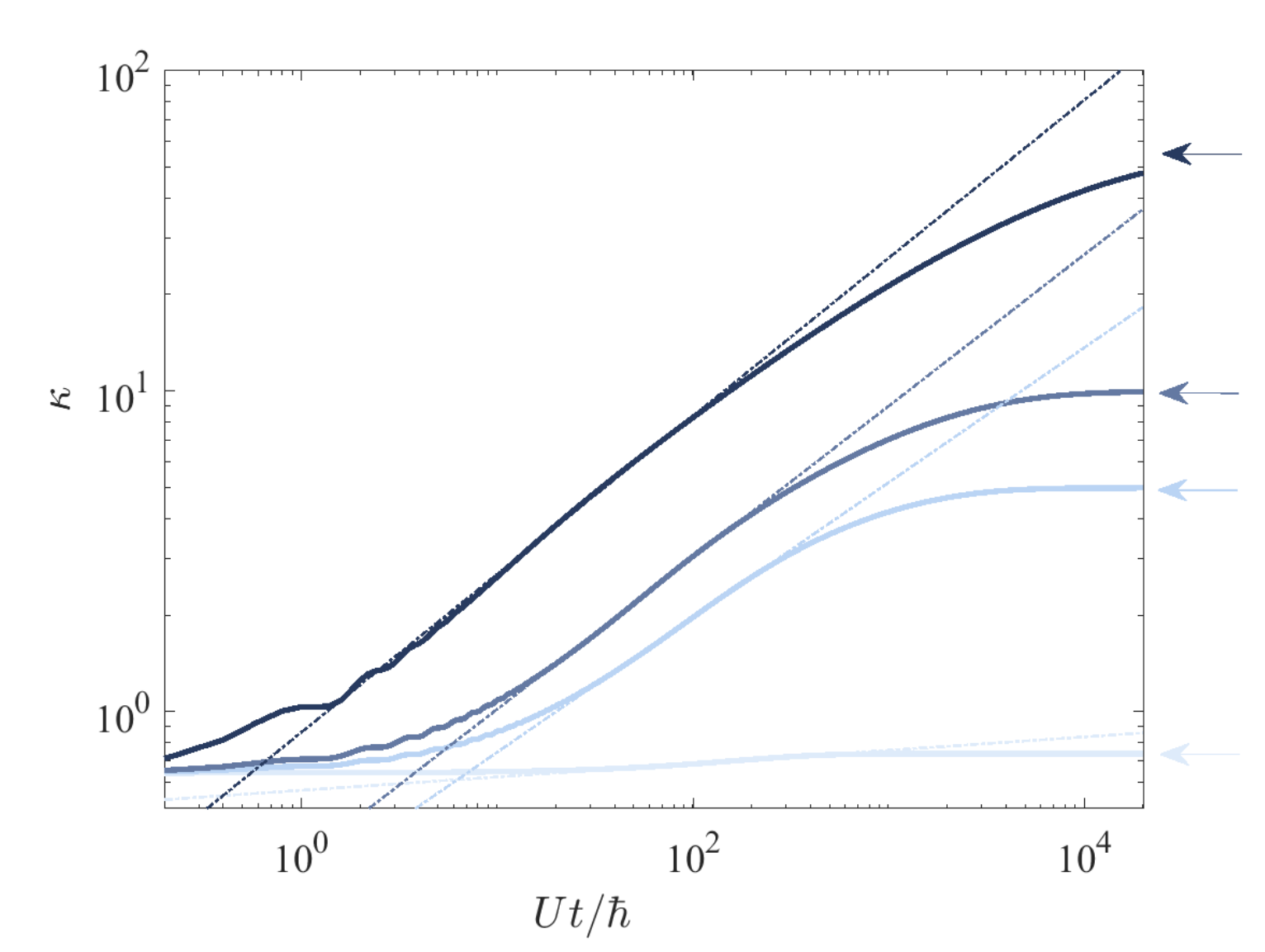}
	\caption{Density fluctuations (solid lines) $\kappa$ vs time $t$, for temperatures (from bottom to top, and from lighter to darker) $T/J = 10, 100, 200,$ and $1200$, with cutoff frequency $\omega_c/J = 500$. Here, the interaction strength is $U/J = 10$ and $N=40$. The dot-dashed lines indicate the algebraic behaviors fitted in the power-law regime, while the arrows point towards the asymptotic values of $\kappa$.}
	\label{fig:Fig3}
\end{figure} 

We now consider temperatures which are low and we plot in Fig. \ref{fig:Fig3} the evolution of the rescaled fluctuations for $U/J=10$, $N=40$, $\omega_c=500J$ and for different temperatures. The various continuous lines represent different temperatures, from $T/J=10$ to $T/J=1200$ from the bottom to the top, while the asymptotic values of the rescaled fluctuations are signaled by arrows. Also in Fig. \ref{fig:Fig3}, it is apparent that the steady-state value of the rescaled fluctuations is different for different temperatures.   
More interestingly, though, we remark that even at low temperatures, there is a power-law regime which can be present for one or two orders of magnitudes of the time evolution. Moreover, the exponent appears to decrease for lower temperatures (as highlighted by the dot-dashed lines which are fits in the power-law regions). 

\begin{figure}[t]
	\centering
	\includegraphics[width=\columnwidth]{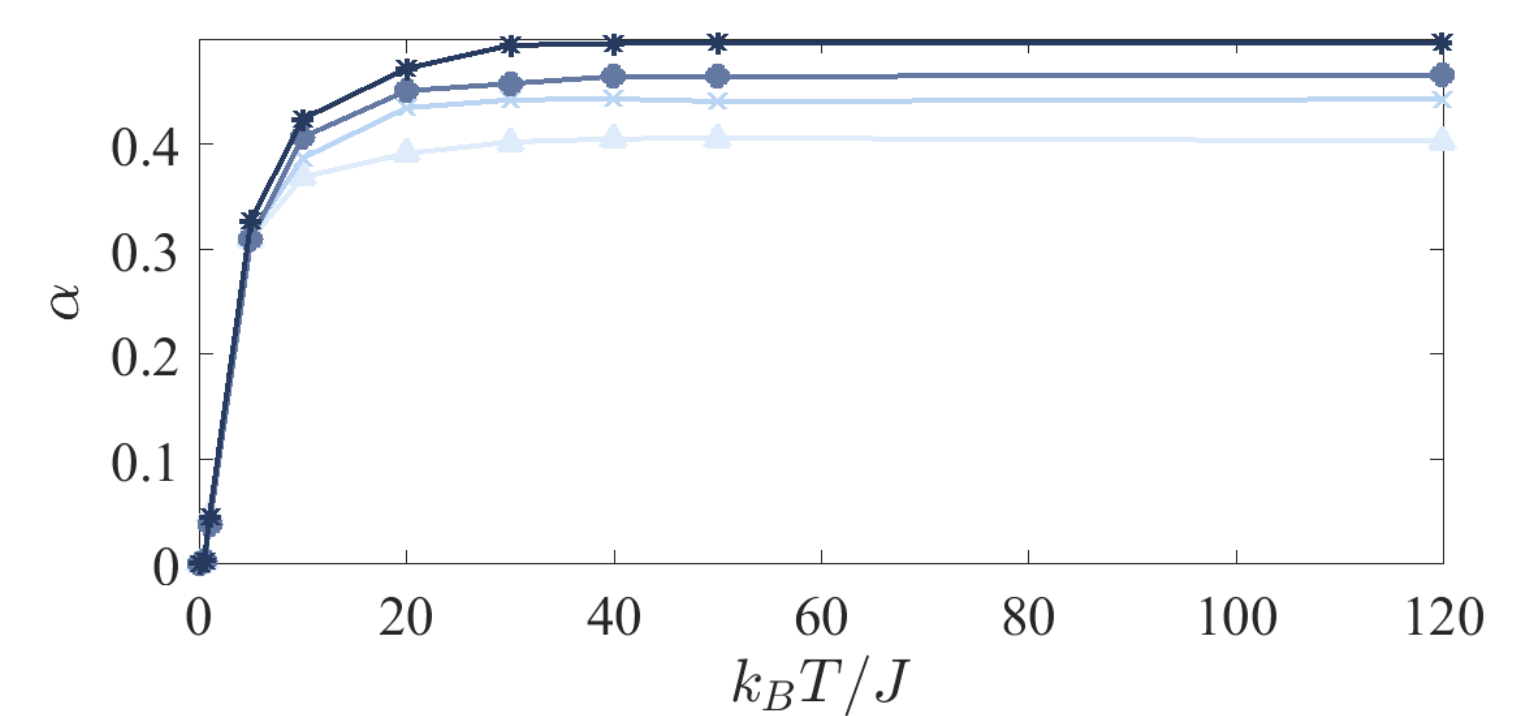}
	\caption{Exponent $\alpha$ of the power-law regime vs temperature $T$ for interaction strengths $U/J=10$, $N=40$, and $\gamma=10^{-2} J$. Different lines, from bottom (light) to top (dark), are for different cutoff frequency $\omega_c/J=10, 50, 100, 1000$. }
	\label{fig:Fig4}
\end{figure}

We thus study the dependence of the power-law exponent as a function of temperature in Fig. \ref{fig:Fig4}. In this figure, the different lines correspond to different frequencies' cutoff, where, more precisely, larger $\omega_c$ are represented by darker lines.  
For large enough $\omega_c$, we observe that as the temperature increases, the exponent of the power law approaches $0.5$, which is the value predicted in Ref. \cite{PolettiKollath2012} for an infinite temperature with local dephasing. However, for lower temperatures, the power-law exponent decreases and, as is naturally expected, for a bath temperature tending to zero, the power-law exponent is $0$. 
For lower $\omega_c$, the overall behavior is similar, in the sense that the exponent is $0$ at $T \rightarrow 0$ and increases for larger temperatures. 
However, the exponent converges, at high temperature, to a value which is smaller than $0.5$. This slower dynamics is related to the fact that for small $\omega_c$, it is necessary to rely on slower and higher-order processes to heat up the interacting system which has large energy gaps.

\section{Conclusions}\label{sec:conclusions} 
We have analyzed the heating dynamics of a strongly interacting bosonic double well in contact with finite-temperature baths via a dephasing interaction. 
For the infinite-temperature local Lindblad studied in the literature, an anomalous diffusive dynamics emerges which results in an algebraic relaxation regime.  
However, for finite temperatures and for baths with small cutoff frequencies $\omega_c$, the duration of the algebraic regime can be much shorter, and the emerging algebraic exponent, if present, can be smaller.  
Hence, the interplay between interaction, kinetic energy, and the bath's properties can result in an even more complex dynamics. While in Ref. \cite{PolettiKollath2012} it was shown that the relaxation could give significant insight into the properties of the spectrum of the system, here we show that such heating dynamics can also give important insight into the baths' properties. We stress that in order to observe algebraic dynamics for longer times in such strongly interacting systems, we had to consider temperatures as high as $T=1200J$, indicating that the infinite-temperature local Lindblad limit, observed experimentally with ultracold atoms \cite{BouganneGerbier2019}, can be difficult to obtain in other setups. 

In the future, we plan to search for a general framework linking the transition rates and the heating dynamics in order to characterize the interplay between the bath spectrum, temperature, and system's level spacing. 
We also plan to gain deeper analytical insights into this much more complex dynamics, as compared to the infinite-temperature case, and to analyze systems beyond the double well, for instance a chain or a higher-dimensional system. Such systems could then be tested in solid-state setups. We note that in systems larger than a double well, the emergence of an even slower dynamics than algebraic decay was predicted \cite{PolettiKollath2013, SciollaKollath2015}, which can be further modified by finite temperature or small $\omega_c$.

\begin{acknowledgments}
 D.P. acknowledges fruitful discussions with R. Fazio and C. Kollath. D.P. acknowledges support from  Ministry of Education of Singapore AcRF MOE Tier-II (Project  No. MOE2018-T2-2-142). The computational work for this article was partially performed on resources of the National Supercomputing Centre, Singapore \cite{nscc}.    
\end{acknowledgments}

\bibliographystyle{apsrev4-1,unsrt}

\begin{thebibliography}{38}
\bibitem [{fn2()}]{fn2}%
  \BibitemOpen
  \href@noop {} {}\bibinfo {note} {For instance, in the power-law regimes which
  we will describe later, $\epsilon \approx 10^{-12}$, and only at earlier
  times may it reach values of, at most, $\epsilon \approx 10^{-6}$. Such small
  values confirm that the description of the time evolution of the particle
  fluctuations, the main focus of our paper, is accurate.}\BibitemShut {Stop}%
\end{thebibliography}

\end{document}